\documentclass[a4paper,11pt]{article}
\pdfoutput=1 

\usepackage{jheppub} 
\usepackage{multirow}
\usepackage{amsmath}
\usepackage{slashed}
\usepackage{amstext,amssymb}
\usepackage{graphicx}
\usepackage{graphicx,wrapfig,lipsum}
\usepackage{xmpmulti}
\usepackage{animate}
\usepackage{epstopdf}
\usepackage{xcolor,colortbl,subfigure}
\usepackage{multimedia}
\usepackage{hyperref}
\usepackage{url}
\usepackage{xspace}
\usepackage{color}
\usepackage{units}
\usepackage{comment}
\newcommand{\mathsym}[1]{{}}

\newcommand{\baz}{\begin{array}{cc}}
\newcommand{\bad}{\begin{array}{ccc}}
\newcommand{\ba}{\begin{array}{c}}
\newcommand{\ea}{\end{array}}
\newcommand{\be}{\begin{equation}}
\newcommand{\ee}{\end{equation}}
\newcommand{\bea}{\begin{eqnarray}}
\newcommand{\eea}{\end{eqnarray}}

\newcommand{\bi}{\begin{itemize}}
\newcommand{\ei}{\end{itemize}}
\newcommand{\bmt}{\begin{pmatrix}}
\newcommand{\emt}{\end{pmatrix}}
\newcommand{\bt}{\begin{tabular}}
\newcommand{\et}{\end{tabular}}

\newcommand{\benu}{\begin{enumerate}}
\newcommand{\eenu}{\end{enumerate}}

\newcommand{\bav}{\begin{array}{cccc}}
\usepackage{wrapfig}
\definecolor{light-gray}{gray}{0.95}
\usepackage{tcolorbox}

\begin{document}

\title{
Charged Lepton Flavour Violating Meson Decays in Seesaw Models}
 \author{Pravesh Chndra Awasthi,} 
 \author{Jai More,}
 \author{Akhila Kumar Pradhan,}
 \author{Kumar Rao,}
 \author{Purushottam Sahu}
\author{and S. Uma Sankar }
\affiliation{Department of Physics, Indian Institute of Technology Bombay, Powai, Maharashtra 400076 India}

\emailAdd{214120016@iitb.ac.in}
\emailAdd{more.physics@gmail.com}
\emailAdd{akhilpradhan@iitb.ac.in}
\emailAdd{kumar.rao@phy.iitb.ac.in}
\emailAdd{purushottam.sahu@iitb.ac.in}
\emailAdd{uma@phy.iitb.ac.in}

\abstract{The occurrence of neutrino oscillations demands the existence of flavour
violation in charged lepton sector. The relation between the branching ratios of
different charged lepton flavour violating (CLFV) decay modes depends on the details of the 
neutrino mass model. In this work, we consider the three types of simple seesaw mechanisms
of neutrino masses and study the correlation between the radiative CLFV decays and the meson CLFV decays. We find that the meson CLFV decay branching ratios are negligibly small in type-II seesaw mechanism whereas they are constrained to be at least three (two) orders of magnitude smaller than the 
radiative CLFV decay branching ratios in the case of type-I (type-III) seesaw mechanism. Thus 
the relationship between these two modes of CLFV decays helps in distinguishing between 
different types of seesaw mechanism. If, the branching ratios of CLFV decays of mesons
are larger than those of radiative CLFV decays, it provides a strong hint that the 
neutrino mass generating mechanism is more complicated than simple seesaw.

}

\maketitle
\flushbottom

\newpage
\section{Introduction}
The discovery of neutrino oscillations~\cite{Cleveland:1998nv, Kamiokande-II:1989hkh, SAGE:1999nng, GALLEX:1998kcz, Super-Kamiokande:1998qwk, SNO:2001kpb, Gajewski:1992iq, Kamiokande-II:1992hns,  Kamiokande:1994sgx, Super-Kamiokande:1998kpq} showed that the neutrinos have masses in the sub-eV range \cite{Esteban:2020cvm,deSalas:2020pgw,Capozzi:2018ubv}. It is a challenge to introduce such tiny masses in the Standard Model (SM). One can trivially generate Dirac masses for neutrinos through standard Higgs mechanism, if the particle content of the SM is extended to include three right-handed (RH) neutrinos. However, this requires  an extreme fine-tuning, ${\mathcal O} \left( 10^{-12} \right)$, of the neutrino Yukawa couplings.
Also, such a mechanism of neutrino mass generation does not lead to any interesting new physics signal.

Flavour oscillations in neutrino sector are possible only if there is
a mixing of different neutrino flavours. Since the left-handed (LH) neutrinos form $SU(2)_L$ doublets
with charged leptons, flavour mixing in neutrino sector necessarily leads
to flavour mixing in charged lepton sector. In particular, neutrino mixing
will lead to charged lepton flavour violating (CLFV) decays such as $\mu \to e \, \gamma$ and $K_L \to \bar{\mu} e$.
If the neutrinos have purely Dirac masses, generated through standard Higgs
mechanism, these processes are Glashow, Iliopoulos, Maiani (GIM) suppressed due to the unitarity of the Pontecorvo, Maki, Nakagawa, Sakata (PMNS) mixing matrix. Their amplitudes will be
of the order $(\Delta m^2/M_W^2) \sim 10^{-24}$, where $\Delta m^2$ is the mass-squared difference of the neutrinos,  leading to branching ratios of
the order of $10^{-50}$ \cite{Cheng:1980qt}. Any other mechanism of neutrino mass generation 
will disrupt this GIM cancellation and can lead to much larger branching 
ratios for CLFV processes \cite{Minkowski:1977sc}. The interplay between different
CLFV decays provides strong clues to the neutrino mass generation mechanism. 

Seesaw mechanism is an innovative method to generate tiny neutrino masses
while avoiding extreme fine-tuning \cite{Minkowski:1977sc,Yanagida:1980xy,Mohapatra:1979ia,GellMann:1980vs}. Typically, this mechanism gives
rise to Majorana masses for light neutrinos. The unique dimension-5 operator, which can be constructed using only the SM fields, is the Weinberg operator \cite{Weinberg:1979sa}. It is
of the form
\begin{equation}
    {\mathcal L}_{d=5} = \frac{1}{2}\frac{C_\Lambda}{\Lambda} \left(\overline{L^c_{L}}\tilde{H}^\ast\right)\left( \tilde{H}^\dagger L_L\right)+ \text{h.c.},
\end{equation}
and it generates Majorana masses for LH neutrinos on electroweak 
spontaneous symmetry breaking. Here, $\Lambda$ is the scale of some 
beyond the Standard Model physics from high energy and the
Weinberg operator is the effective operator coming from this high scale. 
The neutrino mass can be made quite small by choosing either 
$\Lambda$ to be large or $C_\Lambda$ to be small or both. 

It is expected that the Weinberg operator arises due to the exchange
of a heavy particle between neutrino and the Higgs fields. Depending 
on the nature of this exchange particle, seesaw mechanisms are classified
into three types. They are
\begin{itemize}
    \item \underline{Type-I:} The exchange particle is an $SU(2)_L$
    singlet fermion, with hypercharge $Y=0$. 
    \item \underline{Type-II:} The exchange particle is a member of 
    an $SU(2)_L$ triplet of scalars, with hypercharge $Y=2$.
    \item \underline{Type-III:} The exchange particle is a member of
    an $SU(2)_L$ triplet of fermions, with hypercharge $Y=0$.
\end{itemize}
\begin{figure}
    \centering
    \includegraphics[width=0.95\linewidth]{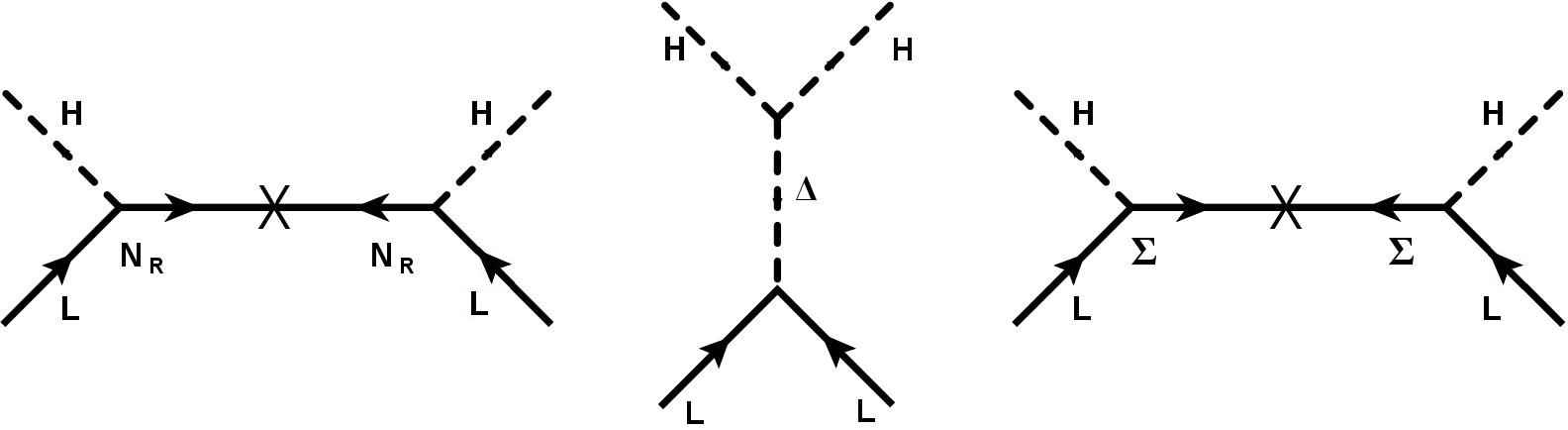}
    \caption{The three realizations of the seesaw mechanism, depending on the nature of the heavy fields exchanged: $SU(2)_L$ singlet fermions (type-I seesaw) on the left,
$SU(2)_L$ triplet scalars (type-II seesaw) in the middle and $SU(2)_L$ triplet fermions (type-III seesaw) on the
right.}
    \label{seesaw type}
\end{figure}
The three types of seesaw are illustrated in figure \ref{seesaw type}.
Even though, the
three types of seesaw give rise to the same form of light neutrino mass
matrix, the BSM couplings of charged leptons in different types are quite
different, which leads to different types of new physics signals. By studying
these NP signals in detail, we can hope to make a distinction between 
different types of seesaw.

There have been a number of detailed studies of CLFV decays of leptons into
a lighter lepton and a photon (which we call radiative CLFV decays) in the three seesaw mechanisms~\cite{Minkowski:1977sc,Bilenky:1977du,Ilakovac:1994kj,Abada:2008ea}. Most of these studies have also considered the CLFV decays of leptons to a set of lighter leptons as well as the $\mu-e$ conversion in the neighborhood of a heavy nucleus~\cite{Dinh:2012bp,Alonso:2012ji}. 
These studies have derived strong upper bounds on the parameters of 
each of the seesaw mechanisms from the above decays. Relatively less 
attention is paid to the CLFV decays of mesons. Comparing the radiative CLFV decays with the CLFV decays of mesons, we note that their amplitudes depend on the same set of flavour violating seesaw parameters. Hence the bounds on
these parameters, derived from the radiative CLFV decays, translate into
bounds on the branching ratios of CLFV decays of mesons~\cite{Gagyi-Palffy:1994xuz}. In other words,
we can obtain a strong constraint relation between the branching ratios of radiative 
CLFV decays and meson CLFV decays. The nature of this relation depends 
on the seesaw mechanism in operation. Hence, a complete information on
all the CLFV decays will help in pinning down the seesaw mechanism.

In this study, we consider the complete set of CLFV meson decays. They
can be purely leptonic, such as $M \to \ell_\beta^+ \ell_\alpha^-$, or 
semi-leptonic, such as $M \to M^\prime \ell_\beta^+ \ell_\alpha^-$ and 
$M \to V \ell_\beta^+ \ell_\alpha^-$. Here $M$ and $M^\prime$ are pseudo-scalar
mesons, $V$ is a vector meson and $\ell_{\alpha}$ and $\ell_{\beta}$ are charged leptons of
different flavours. All these processes involve flavour change in 
quark sector as well as in lepton sector. There is exact GIM suppression
in the quark sector but the corresponding suppression in the lepton 
sector is not exact due to the non-unitarity of the PMNS matrix arising
from the seesaw mechanism. The nature of this non-unitarity depends on 
the type of seesaw mechanism, thus leading to different predictions for
the CLFV decay branching ratios of mesons in different seesaw mechanisms.
The upper bounds on purely leptonic branching ratios are quite strong
because the amplitudes are helicity suppressed. Since this suppression 
is not present for the semi-leptonic decays, we get larger allowed branching
ratios for these decays.

\section{ Seesaw Models: An overview}
\subsection{Type-I seesaw}\label{TI}
In canonical type-I seesaw mechanism 
\cite{Minkowski:1977sc,Yanagida:1980xy,Mohapatra:1979ia,GellMann:1980vs},
one extends the SM by adding $k$ RH neutrino states, $N_R$, to generate tiny neutrino masses.  Since the RH neutrinos are gauge singlets, they can have soft Majorana masses. In addition, they have Yukawa couplings to the LH doublet $L$ and the Higgs doublet $H$. The minimum value of $k$ is 2, so that at least two of the light neutrinos have non-zero masses. The corresponding Lagrangian for Yukawa interaction terms and the soft mass terms, in  type-I seesaw, is
\begin{equation}
 \mathcal{L}_{\rm Type-I} =  -\overline{L} \ Y_D \ \widetilde{H} \ N_{R}
\ - \
{1\over 2} \overline{N^c_R} \ M_R \ N_{R} + \ \text{h.c.},
 \label{eq:type1Lag}
\end{equation}
where $\tilde{H}=i \sigma_2 H^*$. Here $Y_D$ is a $3 \times k$ complex matrix of Yukawa couplings and $M_R$ is a $k \times k$ complex symmetric matrix of Majorana masses. 
Once $H$ gets the vacuum expectation value (\textit{vev}), $\langle H\rangle =v/\sqrt 2$, neutrinos acquire Dirac masses $M_D = Y_D \ v / \sqrt{2}$, which lead to mixing between lepton flavours. The complete set of mass terms in type-I seesaw mechanism is
\begin{eqnarray}
\mathcal{L}_{\rm M} &&= \frac{1}{2} \, \left( \bar{\nu}_{L} \, M_D \, N_{R}  +  \overline{N^c_R} \,M_D^T \, \nu^c_L
 + \ \overline{N^c_R} \ M_R \ N_{R} \right) \ + \ \text{h.c.} \nonumber \\
&&= \frac{1}{2}\begin{pmatrix} \bar{\nu}_L\
\overline{N^c_R}\end{pmatrix} \begin{pmatrix} 0 & M_D \\ M_D^T &
 M_R\end{pmatrix} \begin{pmatrix} {\nu^c_L} \\ N_R\end{pmatrix} + {\rm h.c.}, \\
  &&= \frac{1}{2}\bar{n}_L M n^c_L + {\rm h.c.} \nonumber 
\label{eq:lag_nu_mass}
\end{eqnarray}

The matrix $M$ is a $(3+k) \times (3+k)$ complex symmetric matrix in neutrino flavour basis and it can be transformed into the mass basis 
through a $(3+k) \times (3+k)$ unitary matrix $U$ as
\begin{eqnarray}
 && n_L =  \begin{pmatrix}
  \nu_{L} \\ N^c_{R}  
\end{pmatrix}
     = U \begin{pmatrix} \nu_{i L\,} \\ N^c_{k R\,} \end{pmatrix}
     = \begin{pmatrix} U_{\nu\nu} & U_{\nu N} \\ U_{N\nu} & U_{NN} \end{pmatrix}   
          \begin{pmatrix} \nu_{i L\,} \\ N^c_{k R\,} \end{pmatrix}, 
          \nonumber \\
   && n^c_L = \begin{pmatrix}
  \nu_{L\,}^c \\ N_{ R}  
\end{pmatrix}
     = U^* \begin{pmatrix} \nu^c_{i L\,} \\ N_{k R\,} \end{pmatrix} 
     = \begin{pmatrix} U_{\nu\nu}^* & U_{\nu N}^* \\ U_{N\nu}^* & U_{NN}^* \end{pmatrix}   
          \begin{pmatrix} \nu_{i L\,}^c \\ N^c_{k R\,} \end{pmatrix}.      
\end{eqnarray}
We transform $M$ from the flavour basis to the mass basis in two steps. First, we transform it to a block diagonal form through a unitary matrix $W$ as 
\begin{eqnarray}
W^T M W &=&
W^T
\left(
  \begin{array}{cc}
    0 & M_D \\
    M_{D}^{T} & M_R \\
  \end{array}
\right) W =
\left(
  \begin{array}{cc}
    m_\nu & 0 \\
    0 & M_N  \\
  \end{array}
\right) = M_I,
\label{eq:type1NuMixMatrix}
\end{eqnarray}
where $m_\nu$ is a $3 \times 3$ complex symmetric matrix of light neutrino Majorana masses and $M_N$ is a $k \times k$ complex symmetric matrix of heavy neutrino Majorana masses. We need $M_D \ll M_R$ for seesaw mechanism to generate tiny light neutrino masses without fine tuning. The matrices $m_\nu$ and $M_N$ are transformed to diagonal form through the  unitary matrix 
\begin{equation} 
U_I= \left(
  \begin{array}{cc}
    U_\nu & 0 \\
    0 & U_N  \\
  \end{array}
\right)
\end{equation} 
such that 
\begin{eqnarray}
U_I^T M_I U_I &=&
\left(
  \begin{array}{cc}
    m^{diag}_\nu & 0 \\
    0 & M^{diag}_N  \\
  \end{array}
\right) = M^{diag}=diag(m_1,m_2,m_3,M_1,...M_k).
\label{eq:type1NuMixMatrix}
\end{eqnarray}
Now the complete diagonalising matrix is $U = W \, U_I$. Using the formalism of ref.~\cite{Korner:1992zk,Grimus:2000vj}, the matrix $W$ can be written as 
\begin{eqnarray}\label{eq:Wmatrix}
W &\equiv& 
  \begin{pmatrix} \sqrt{\mathbb{I} - RR^\dagger} &
R \\ -R^\dagger & \sqrt{\mathbb{I} - R^\dagger R}\end{pmatrix} \simeq \begin{pmatrix} \mathbb{I} - \frac{1}{2}RR^\dagger &
R \\ -R^\dagger & \mathbb{I} - \frac{1}{2}R^\dagger R\end{pmatrix} 
\end{eqnarray}
where, $R$ is a $3\times k$ matrix, given by $M_D^* M_R^{-1} $.
With this form of $W$, the matrix
$U$ takes the form
\begin{equation}\label{U in type-I}
    U=\begin{pmatrix} U_{\nu\nu} & U_{\nu N} \\ U_{N\nu} & U_{NN} \end{pmatrix} = \begin{pmatrix} (\mathbb{I} - \frac{1}{2}RR^\dagger) \, U_\nu &
R U_N \\ -R^\dagger U_\nu & (\mathbb{I} - \frac{1}{2}R^\dagger R) \, U_N \end{pmatrix}.  
\end{equation}
The PNMS matrix which connects the neutrino flavor state to the light mass eigenstates is no longer unitary because of light-heavy neutrino mixing arising through $R$
 \begin{equation*}
     U_{PMNS}= \left(\mathbb{I} - \frac{1}{2}RR^\dagger \right) \, U_\nu = (\mathbb{I} + \eta ) \, U_\nu. \,
 \end{equation*}
 
 For the matrix $R$, the first index is a flavor index and the second index is heavy neutrino mass index, which implies that the matrix $\eta \sim R R^{\dagger} = (R)_{\alpha k} (R^{\dagger})_{k \beta} = (RR^{\dagger})_{\alpha \beta}$ has two flavor indices. $U_\nu$ has its first index as flavor and second index as light neutrino mass eigenstate. Hence the product $(\mathbb{I} - \frac{1}{2}RR^\dagger) \, U_\nu $ by definition has first index to be flavor and second index to be light neutrino mass $(\mathbb{I} - \frac{1}{2}RR^\dagger)_{\alpha \beta} \, (U_\nu)_{\beta i} $.
 The elements of the Hermitian matrix $\eta$ can be constrained by utilizing existing neutrino oscillation data as well as data from electroweak processes. Such global constraints on the non-unitary mixing can be found in ref.~\cite{Fernandez-Martinez:2016lgt,Blennow:2023mqx,Fernandez-Martinez:2024bxg},
\begin{equation}
    |\eta_{\alpha \beta}|\leq\begin{pmatrix}
     1.3\times10^{-3} &1.2\times10^{-5} &1.4\times10^{-3} \\
     1.2\times10^{-5} &2.2\times10^{-4} &6\times10^{-4} &\\
     1.4\times10^{-3} &6\times10^{-4} &2.8\times10^{-3} 
    \end{pmatrix}.
\end{equation}

In the limit $M_D\ll M_R$, the light $(m_\nu)$ and heavy $(M_N)$ neutrino mass matrices are
\begin{eqnarray}\label{mlight}
m^I_\nu\approx  -  M_D M^{-1}_R M_D^T \quad\text{and}\quad  M_N\approx M_R.
\end{eqnarray}
Without loss of generality, one can assume $M_R$ is diagonal which means $U_N = I$. 
The neutrino flavour eigenstate can be written as
\begin{eqnarray}
 \nu_{\alpha} = \left[(\mathbb{I}+\eta) \, U_\nu \right]_{\alpha \, i}\, \nu_i  + R_{\alpha k} \, N_k .
\end{eqnarray}


In the following, we will work in the basis where the charged lepton mass matrix is diagonal. The charged current (CC) and neutral current (NC) weak interaction couplings involving the light Majorana neutrinos \(\nu_i\), which have definite masses \(m_i\), are given by:
\begin{equation}
\begin{split}
    \mathcal{L}_{C C}^\nu &=-\frac{g}{2 \sqrt{2}} \bar{\ell}_{\alpha} \gamma_\mu (1-\gamma_5) \nu_{\alpha} W^\mu+\text {h.c.}\\
&=-\frac{g}{2 \sqrt{2}} \bar{\ell}_{\alpha} \gamma_\mu (1-\gamma_5) \left[(1 +\eta) \, U_\nu \right]_{\alpha \, i} \nu_i\,  W^\mu+\text {h.c.}, \\
\end{split}
\end{equation}
\begin{equation}
    \begin{split}
        \mathcal{L}_{N C}^\nu &=-\frac{g}{2 \cos \theta_W} \bar{\nu}_{\alpha L}
        \gamma_\mu \nu_{\alpha L} Z^\mu\\
       & =-\frac{g}{2 \cos{\theta_W}} \bar{\nu_i}_L \gamma_\mu\left(U_\nu^{\dagger}(1+\eta+\eta^{\dagger}) U_\nu\right)_{i j} \nu_{j L} Z^\mu .
    \end{split}
\end{equation}
The charged current and the neutral current interactions of the heavy Majorana fields $N_k$ with $W^{ \pm}$and $Z^0$ read:
\begin{equation}
\begin{aligned}
\mathcal{L}_{C C}^N & =-\frac{g}{2 \sqrt{2}} \bar{\ell}_\alpha\, \gamma_\mu R_{\alpha k}\left(1-\gamma_5\right) N_k W^\mu+\text { h.c. } \\
\mathcal{L}_{N C}^N & =-\frac{g}{2 \cos 
\theta_W} \bar{\nu}_{\alpha L} \gamma_\mu R_{\alpha k} N_{k L} \, Z^\mu+\text { h.c. }
\end{aligned}
\end{equation}
\subsection{Type-II seesaw}
To generate light neutrino masses through type-II seesaw mechanism \cite{Magg:1980ut,Lazarides:1980nt,Schechter:1980gr}, an $SU(2)_L$ scalar triplet field $(\Delta)$ with hypercharge $Y=2$, is added to the SM. This field couples to SM lepton doublets and their charge conjugates via Yukawa interactions.  The $\Delta$ field can be expressed in its $2\times2$ matrix representation as
\begin{equation}
\Delta=\left(\begin{array}{cc}
\Delta^{+} / \sqrt{2} & \Delta^{++} \\
\Delta^0 & -\Delta^{+} \sqrt{2}
\end{array}\right)
\end{equation}
where $\Delta^0, \Delta^{+}$and $\Delta^{++}$ are neutral, singly and doubly charged components. The terms in the Lagrangian corresponding to the $\Delta$ field are
\begin{equation}
-\mathcal{L}_{\text {Type- II }}=\left(Y_{\Delta} \,L^T\,C i\sigma_2 \,\Delta \, L+\mu_{\Delta} H^T  i\sigma_2\,\Delta^{\dagger} \,H+\text { h.c. }\right)
+M_{\Delta}^2 \text{Tr}\left(\Delta^{\dagger} \Delta\right),
\end{equation}
where $Y_{\Delta}$ is the complex Yukawa coupling matrix, $C$ is the charge conjugation matrix, $\mu_{\Delta}$ is the dimensionful coupling constant connected to the lepton number violating term and $M_{\Delta}$ stands for the mass of the scalar triplet. When the neutral component of $\Delta$ acquires nonzero \textit{vev}  $\left(\left\langle\Delta^0\right\rangle=v_{\Delta}\right)$, a Majorana mass matrix for the LH neutrinos is generated 
\begin{equation}
m_\nu^{I I}=2 \, Y_{\Delta} \, v_{\Delta} \text { with } v_{\Delta}=\frac{\mu_{\Delta} \, v^2}{M_{\Delta}^2}.
\label{typeII}
\end{equation}
The mass scale $M_\Delta$ is expected to be much larger than the electroweak scale $v$ because no charged or doubly charged scalars have been observed. This in turn makes $\mu_\Delta$ quite small, leading to small LH neutrino masses.

\subsection{Type-III seesaw}
It is also possible to generate small Majorana masses for LH neutrinos by adding $SU(2)_L$ fermionic triplet $(\vec{\Sigma}=\Sigma^1,\Sigma^2,\Sigma^3)$ with hypercharge zero \cite{Foot:1988aq,Ma:1998dn} to the SM. At least two such triplets are needed in order to have two non-vanishing neutrino masses. The three components of the field $\vec{\Sigma}$ can be written in matrix form as 
\begin{equation}
\Sigma_k =\left(\begin{array}{cc}
\Sigma^0_k / \sqrt{2} & \Sigma^{+}_k \\
\Sigma^{-}_k & -\Sigma^0_k \sqrt{2}
\end{array}\right),
\end{equation}
where 
\begin{equation*}
    \Sigma^{\pm}_k\equiv \frac{\Sigma^1_k\mp i\Sigma^2_k}{\sqrt{2}} , \,\,\,\, \Sigma^0_k\equiv\Sigma^3_k, \, \, (i = 1,2).
\end{equation*}
The Majorana mass terms of the triplets $\Sigma_i$ are gauge invariant. The mass and Yukawa interaction terms of these $\Sigma_k$ fields are 
\begin{equation}
-\mathcal{L}_{\text {Type- III }}=\frac{1}{2} \text{Tr} \overline{\Sigma}_k (M_{\Sigma})_{k \ell} \Sigma^c_\ell +
\sqrt{2} \tilde{H}^{\dagger} \overline{\Sigma}_k (Y_{\Sigma})_{k \alpha} L_{\alpha L}+ \text { h.c. }
\end{equation}

The Majorana mass matrix $M_{\Sigma}$ can be assumed to be real and diagonal without loss of generality. The Yukawa coupling matrix $(Y_{\Sigma})_{k \alpha}$ is a $k \times 3$ complex matrix, which leads to mixing between the SM leptons and the triplet fermions. 
In order to study this mixing, it is convenient to define the charged Dirac spinor $\Psi\equiv \Sigma^{+ c}_R +\Sigma^-_R$ which mixes with charged leptons. The neutral component of the triplet $\Sigma^0$ mixes with neutrinos.
The lepton mass term in the Lagrangian \cite{Abada:2008ea} (omitting the generation index $k$ of triplets)
\begin{equation}
    \mathcal{L}_{\rm M} =- \left(\overline{l}_L \,\,\,\overline{\Psi}_L\right)\left(\begin{array}{cc}
m_l & Y_{\Sigma}^{\dagger} v \\
0 & M_{\Sigma}
\end{array}\right)\binom{l_R}{\Psi_R}-\frac{1}{2}\left(\overline{\nu^c_L} \,\,\,\overline{\Sigma^0}\right)\left(\begin{array}{cc}
0 & M_D \\
M_D^T & M_{\Sigma}
\end{array}\right)\binom{\nu_L}{\Sigma^{0 c}},
\end{equation}
where $M_D=Y_{\Sigma}^T v / \sqrt{2}$. The neutral fermion mass matrix has the same form as in
the type-I seesaw case. Using the procedure described in subsection \ref{TI}, this matrix is 
first put in block diagonal form
\begin{equation}
\left(\begin{array}{cc}
m_\nu^{III} & 0 \\
0 & M_{\Sigma} \end{array}\right),
\end{equation}
where $m_\nu^{III} = - M_D M_{\Sigma}^{-1} M_D^T$ and then fully diagonalized. The diagonalizing
matrix $U$ has the same form as the one given in eq.~\eqref{U in type-I} and is given by
\begin{equation}
U=\begin{pmatrix}
            (1+\eta)U_{\nu}& R\\
            - R^\dagger U_\nu & 1+\eta^{\prime}
        \end{pmatrix},
\end{equation}
where $U_\nu$ is the diagonalizing matrix of $m_\nu^{III}$, $R = M_D^* M_\Sigma^{-1}$, $\eta = -R R^\dagger/2$ and $\eta^\prime = 
-R^\dagger R/2$. 
The new feature in type-III seesaw is that the mass matrix for charged leptons is a general complex matrix, which is diagonalized by a bi-unitary transformation with the two unitary matrices $U_L$
and $U_R$. To the order $\mathcal{O}\left([(vY_{\Sigma}, m_l)/M_{\Sigma}]^2\right)$,
these matrices are given by 
\begin{equation}
    \begin{split}
          & U_L=\begin{pmatrix}
            1+2\eta & \sqrt{2} R\\
           - \sqrt{2} R^\dagger & 1+2\eta^{\prime}
        \end{pmatrix}\\
        & U_R=\begin{pmatrix}
            1 & \sqrt{2} m_l R  M^{-1}_\Sigma \\
           - \sqrt{2} M^{-1}_\Sigma R^\dagger m_l & 1
        \end{pmatrix}.
        \end{split}
\end{equation}

As in the case of type-I seesaw, the PMNS matrix is related to the unitary matrix $U_\nu$ as 
 \begin{equation*}
     U_{PMNS} = (\mathbb{I} + \eta) \, U_\nu.
 \end{equation*}
Hence, the PMNS matrix in non-unitary in type-III seesaw also.  
The lepton neutral current couplings in type-III seesaw are given by 
\begin{equation}
\mathcal{L}_{NC}^\ell =\frac{g}{2 \cos 
\theta_W} \bar{\ell}_{\alpha}\, \gamma_\mu \left[\left\{ P_L(1-2\cos^2 \theta_W) + P_R \,\left(2\sin^2 \theta_W\right)\right\} \delta_{\alpha \beta} + 4 P_L \, \eta_{\alpha \beta} \right]\,\ell_{\beta}\, Z^\mu+\text { h.c. },   
\end{equation}
where $P_L$ and $P_R$ are chiral projection operators. For $\alpha\neq\beta$, the last term in the above equation leads to 
\begin{equation}
    \mathcal{L}_{FCNC}^\ell = - \frac{g}{2 \cos 
\theta_W} \left(  R_{\alpha k} R^\dagger_{k \beta} \right) \,\bar{\ell}_{\alpha}\,  \gamma_\mu \left(1 - \gamma_5 \right) \ell_{\beta}\, Z^\mu+\text { h.c.}, 
\end{equation}
where we have substituted $\eta = - RR^\dagger/2$.
Unlike type-I see-saw, type-III seesaw contains flavour changing neutral currents (FCNC) in the charged lepton sector.

\section{Radiative $\mu\rightarrow e \, \gamma$ Decay}\label{Raddecay}
For type-I seesaw, the branching ratio for the radiative CLFV decay $\mu \rightarrow e \, \gamma$ with $\nu_i$ (with mass $m_i$) and $N_k$ (with mass $M_k$) exchange can be written as \cite{Ibarra:2011xn} 
 \begin{equation}
    Br \left(\mu \to e \, \gamma \right)=\frac{3\alpha_{em}}{32\pi}|T|^2.
\end{equation}
The amplitude $T$, for this radiative decay, is given by
\begin{eqnarray}
T \cong  \sum_k  R_{e k} R^{\ast}_{\mu k} \left[f(x_k)-f(0)\right]   . 
\end{eqnarray}
The function $f(x_k)$, arising from loop integration, has the form
\begin{equation*}
    f(x_k)=\frac{10-43x_k+78x_k^2-49x_k^3+4x_k^4+18 x_k^3\log (x_k)}{3(x_k-1)^4},
\end{equation*}
where, $x_k=(M_k/M_W)^2$.
We work in the limit where $k=2$, which is the minimum number of RH neutrinos required to generate two non-vanishing neutrino masses. We consider the case where the two heavy RH neutrinos are nearly degenerate \textit{i.e.} $M_2 = M_1 (1+z) \,, z \ll 1$ \cite{Ibarra:2011xn}. In this approximation, 
\begin{eqnarray}
 \sum_k  R_{e k} R^{\ast}_{\mu k}f(x_k) & = & f(x_1) R_{e 1} R^{*}_{\mu 1} + f(x_2) R_{e 2} R^{*}_{\mu 2} \cong f(x_1) \left(R_{e 1} R^{*}_{\mu 1} + R_{e 2} R^{*}_{\mu 2}\right), \nonumber  \\
 T & \cong &  \left(R_{e 1} R^{*}_{\mu 1} + R_{e 2} R^{*}_{\mu 2}\right) \left[ f(x_1)-f(0) \right].
\end{eqnarray} 
As shown in \cite{Ibarra:2011xn}, the elements of the matrix $R$ obey the relation
\begin{equation}
    R_{\alpha 2} = \pm i \frac{1}{\sqrt{1+z}} R_{\alpha 1} \, \, {\rm for} \, \, \alpha = e, \mu, \tau, 
\end{equation}
in the limit $z \ll 1$. Hence, the amplitude is simplified to
\begin{equation}
    T \cong \frac{2 + z}{1+z} R_{e1} R_{\mu 1}^* \left[ f(x_1) - f(0) \right]. 
\end{equation}
Similar expressions also hold for the other radiative CLFV decays $\tau \to e \, \gamma$ and $\tau \to \mu \, \gamma$. 

\begin{table}[h!]
\centering
\begin{tabular}{|c|c|c|c|}
\hline \hline
 LFV Decays  & Present Bound  & Near Future Sensitivity \\
\hline
$Br \left(\mu \to {e\gamma} \right)$                 & $42 \times 10^{-14}$\cite{MEG:2016leq}          & $6\times 10^{-14}$\cite{MEGII:2018kmf}  \\[2mm]
\hline
$Br \left(\tau \to {e\gamma} \right)$   & $33\times 10^{-9}$\cite{BaBar:2009hkt}        &$3\times10^{-9}$\cite{Belle-II:2018jsg} \\[2mm]
\hline
$Br \left(\tau \to {\mu\gamma} \right)$   & $42   \times 10^{-9}$\cite{Belle:2021ysv}       & $2.7 \times 10^{-9}$\cite{Belle-II:2018jsg} \\[2mm]
\hline
\end{tabular}
 \caption{Branching ratios for different radiative CLFV decays, their present experimental bounds and future sensitivity values.}
\label{lfv-expt-bound}
\end{table}

Using the present bounds given in Table \ref{lfv-expt-bound} for the radiative CLFV decays
$\ell_\beta \to \ell_\alpha \, \gamma $, we can set a limit on the product $| R_{\alpha 1} R^{\ast}_{\beta 1}|$ for different masses of RH neutrinos. They are shown in Table \ref{RRst-bounds-T1}, for the two values of heavy RH neutrino mass $M_1 = 100$ GeV and $M_1 = 1$ TeV. The bound on the CLFV parameter
$| R_{\alpha 1} R^{\ast}_{\beta 1}|$ is essentially independent of the heavy 
neutrino mass $M_1$. As described in subsection~\ref{TI}, the matrix $R$ is given by
$R = M_D^\ast \, M_R^{-1}$. Since we assumed the matrix $M_R$ to be 
diagonal with nearly equal heavy neutrino masses, the elements of
$R$ should scale as $(M_1)^{-1}$. The matrix $M_D$ is usually 
parametrized in Casas-Ibarra form~\cite{Casas:2001sr}, which contains 
a number of unknown and unphysical parameters. By choosing these 
parameters appropriately, it is possible to scale elements of $M_D$
in such a way that the $R_{\alpha k}$ have only a very mild 
dependence on the value of $M_1$~\cite{Ibarra:2011xn}. 

\begin{table}[h!]
\centering
\begin{tabular}{|c|c|c|c|}
\hline \hline
 Type-I   & $M_1 = 100$ GeV      & $M_1 = 1$ TeV \\
\hline
 $| R_{e 1} R^{\ast}_{\mu 1}|$ &  $3.43\times 10^{-5}$  & $1.17\times 10^{-5}$     \\[2mm]
\hline
$| R_{e 1} R^{\ast}_{\tau 1}|$    & $9.62\times 10^{-3}$        & $3.28\times10^{-3}$ \\[2mm]
\hline
$| R_{\tau 1} R^{\ast}_{\mu 1}|$   & $11.1\times10^{-3}$       & $3.79 \times 10^{-3}$ \\[2mm]
\hline
\end{tabular}
\caption{ Bounds on the product $| R_{\alpha 1} R^{\ast}_{\beta 1}|$ in type-I seesaw for $z=10^{-3}$.}
\label{RRst-bounds-T1}
\end{table}

In type-II seesaw, the radiative CLFV decays are mediated by the charged and
doubly charged components of the scalar triplet. These particles do not couple
to quarks and hence play no role in the CLFV decays of mesons. Hence, we do not
discuss the limits imposed by the radiative CLFV decays on the seesaw mixings 
in type-II seesaw.

 Similarly, for type-III seesaw, we can write the branching ratio for $\mu \rightarrow e \,\gamma$ with $\nu_i$ (with mass $m_i$) and $\Sigma_k$ (with mass $M_k$) exchange as \cite{Abada:2008ea}
  \begin{equation}
    Br \left(\mu \to e \, \gamma \right)=\frac{3\alpha_{em}}{32\pi}|T|^2
\end{equation}
The amplitude $T$ is given by
\begin{eqnarray}
T \cong \sum_k  R_{e k} R^{\ast}_{\mu k} \left[-2.23 + A(x_k) + B(y_k) + C(z_k)\right] ,   
\end{eqnarray}
where $x_k=(M_k/M_W)^2$, $y_k=(M_k/M_Z)^2$ and $z_k=(M_k/M_H)^2$ and $M_H$ is the mass of the
Higgs boson. The loop functions are given by 
\begin{equation}
    \begin{split}
        & A(x_k)=\frac{-30+153 x_k-198 x_k^2+75 x_k^3 +18(4-3 x_k)x_k^2\log(x_k)}{3(x_k -1)^4},\\
        & B(y_k)=\frac{33-18 y_k-45 y_k^2+30 y_k^3 +18(4-3 y_k)y_k\log(y_k)}{3(y_k -1)^4},\\
        & C(z_k)=\frac{-7+12 z_k+3 z_k^2-8 z_k^3 +6(3 z_k-2)z_k\log(z_k)}{3(z_k -1)^4}.
    \end{split}
\end{equation}
We work in the approximation of near-degenerate limit of neutral triplet fermions $(M_2 = M_1 (1+z), z \ll 1$. As in the case of type-I seesaw, the bounds on the product $|R_{\alpha 1} R^*_{\beta 1}|$ are calculated from the present upper bounds on the branching ratios of the radiative CLFV decays $\ell_\beta \to \ell_\alpha \, \gamma$. These are listed in Table~\ref{RRst-bounds-T3}, for two values of triplet fermion masses, $M_1 = 100$ GeV and $M_1 = 1$ TeV. 
\begin{table}[h!]\label{RRst-bounds-T3}
\centering
\begin{tabular}{|c|c|c|c|}
\hline 
 Type-III   & $M_1 = 100$ GeV      & $M_1= 1$ TeV \\
\hline \hline
 $| R_{e 1} R^{\ast}_{\mu 1}|$ &  $3.69\times 10^{-6}$  & $5.39\times 10^{-6}$     \\[2mm]
\hline
$| R_{e 1} R^{\ast}_{\tau 1}|$    & $1.03\times 10^{-3}$        &  $1.51\times10^{-3}$ \\[2mm]
\hline
$| R_{\tau 1} R^{\ast}_{\mu 1}|$   & $1.19   \times 10^{-3}$       & $1.74 \times 10^{-3}$ \\[2mm]
\hline
\end{tabular}
\caption{Bounds on the product $| R_{\alpha 1} R^{\ast}_{\beta 1}|$ in type-III seesaw for $z=10^{-3}$. }
\end{table}

\begin{figure}
    \centering
    \includegraphics[width=0.8\linewidth]{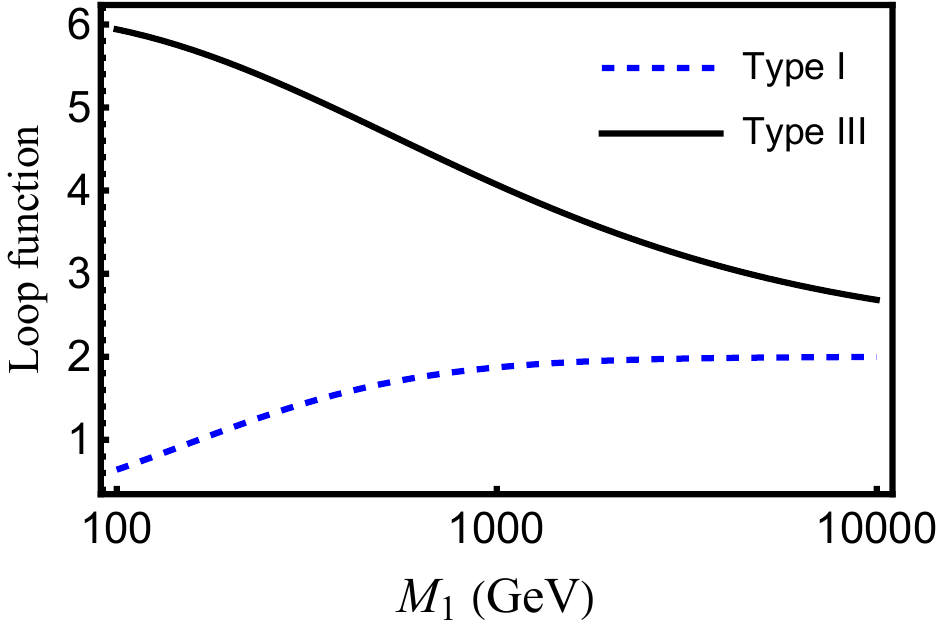}
    \caption{Plots of the loop functions in $\mu \to e\,\gamma$ decay vs heavy neutral fermion mass for type-I (blue/dashed line) and type-III (black/solid line) seesaw }
    \label{figloopfn}
\end{figure}

In both type-I and type-III seesaw mechanisms, the radiative CLFV decay amplitude is a product of the light-heavy mixing matrix elements $R$ and the loop functions. The latter are smooth functions of the heavy neutral fermion mass scale $M_1$. They are plotted in figure~\ref{figloopfn} as functions of $M_1$ for values of $M_1$ in the range $(100,10000)$ GeV. The loop function for type-III seesaw is larger than that for type-I seesaw. To saturate the experimental upper bound on the branching ratios of the radiative CLFV decays, the elements of $R$ in type-III seesaw need to be correspondingly smaller than those for type-I seesaw.

The matrix $R$ is given by $R = M_D^\ast M_R^{-1}$ in type-I seesaw and $R = M_D^\ast M_\Sigma^{-1}$ in type-III seesaw. In both cases, the Dirac mass matrix $M_D$ can be parametrized in the Casas-Ibarra form~\cite{Casas:2001sr}. For normal mass ordering of light neutrinos, the elements of $R$ can be written as~\cite{Dinh:2012bp}
\begin{equation}
    |R_{\ell \,1}| = \left(\frac{y\,v}{\sqrt{2}} \right) \frac{1}{M_1} \sqrt{\frac{m_3}{m_2+m_3}}
    \left|U_{\ell \,3}\, + i \sqrt{m_2/m_3} \, U_{\ell\,2} \right|
    \label{defRl1}
\end{equation}
with $|R_{\ell\, 2}| \approx |R_{\ell\, 1}|$ for $M_2 \approx M_1$.
In the above equation, $(y\,v/\sqrt{2})$ is the maximum eigenvalue of $M_D$, $U_{\ell \, (2,3)}$ are elements of the PMNS matrix and $m_2$ and $m_3$ are the second and third light neutrino masses. For particular values of the phases of the PMNS matrix, it is possible for the elements of $R$ to be vanishingly small. However, these cancellations are independent of the value of $M_1$ and depend only on the phases.

In literature, two types of CLFV processes were studied extensively: radiative decays and $\mu-e$ conversion. For the radiative decays, the loop functions are smooth functions of the seesaw mass scale $M_1$ whereas the loop function in $\mu-e$ conversion has cancellations for some particular values of $M_1$. Hence, the plot of $Br(\mu \to e\,\gamma)$ is a smooth function but the ratio
of $\mu-e$ conversion rate to $Br(\mu \to e \, \gamma)$ shows sharp dips, as can be seen in ref.~\cite{Dinh:2012bp}.


\section{CLFV meson decays in seesaw models}\label{Sec-4}

In this section, we consider the effective Hamiltonian for the transition 
$q_1 \to q_2 \,\ell_\beta^+ \ell_\alpha^-$, where $q_1$ and $q_2$ are two different quarks of charge $-1/3$. The quark flavour change occurs at loop level, due to CC interactions at second order, leading to the quark currents of $(V-A)$ structure. Hence, the generic four Fermi Hamiltonian to describe the above transitions can be written as
\begin{equation}
\label{EFT}
    H_{eff} = \frac{4 G_F}{\sqrt{2}}\sum_{j=u,c,t} V^*_{j\,q_1} V_{j \, q_2} \, [C_9\, O_9 + C_{10}\,O_{10}],
\end{equation}
where $V^*_{j\,q_1} V_{j \, q_2}$ is the product of the Cabibbo-Kobayashi-Maskawa (CKM) matrix elements. 
The operators $O_9$ and $O_{10}$ have Wilson coefficients $C_9$ and $C_{10}$ respectively.
These operators are defined as 
\begin{equation}
\begin{split}
     & O_9=\frac{e^2}{(4\pi)^2}[\bar{q}_2\,\gamma^{\mu}P_L q_1] [\bar{\ell}_\alpha\,\gamma_{\mu}\,\ell_\beta] 
    =\frac{\alpha}{8\pi}[\bar{q}_2\,\gamma^{\mu}\left(1-\gamma_{5}\right) q_1] [\bar{\ell}_\alpha\,\gamma_{\mu}\,\ell_\beta], \nonumber \\
&O_{10}=\frac{e^2}{(4\pi)^2}[\bar{q}_2\gamma^{\mu}P_L q_1] [\bar{\ell}_\alpha\gamma_{\mu}\gamma_5\, \,\ell_\beta]
    =\frac{\alpha}{8\pi}[\bar{q}_2\gamma^{\mu}\left(1-\gamma_{5}\right) q_1] [\bar{\ell}_\alpha\gamma_{\mu}\gamma_5\, \,\ell_\beta]. \nonumber
\end{split}
\end{equation}
In the expressions for $C_9$ and $C_{10}$, the GIM suppression in the quark sector is operative, which makes these coefficients to be 
proportional to $m_j^2$. The product 
$\left(V_{j q_1}^* V_{j q_2} m_j^2\right)$ is largest for
$j = t$, for all three combinations
$(\bar{q}_1 \, q_2) = (\bar{s} \, d), (\bar{b} \, d)$ and $(\bar{b} \, s)$. Hence, we can drop the terms $j = c$ and $j = u$, in the summation in eq.~\eqref{EFT}.

\begin{figure}
     \centering
     \includegraphics[width=0.95\linewidth]{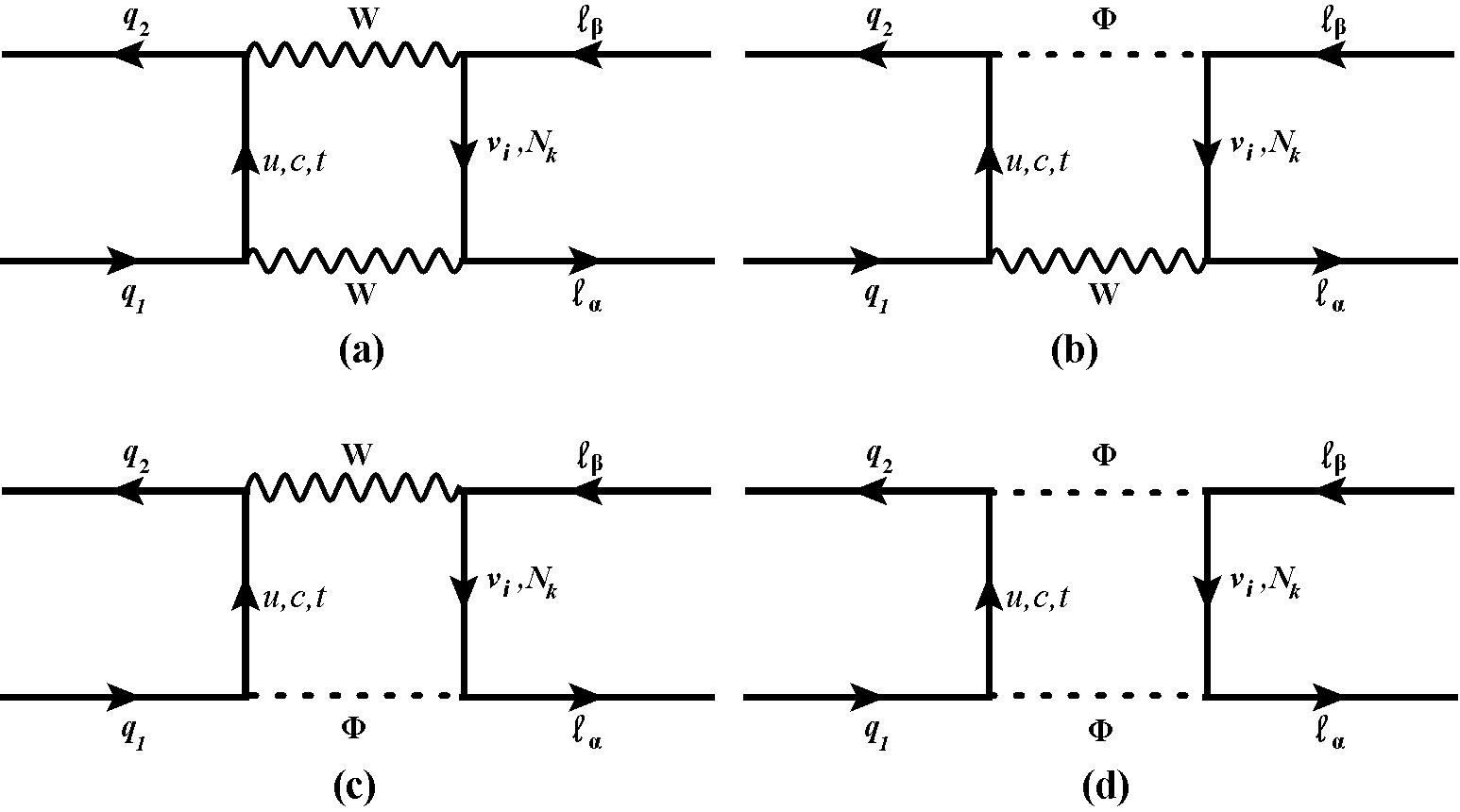}
     \caption{Diagrams contributing to $q_1 \to q_2 \ell_\beta^+ \ell_\alpha^- $ CLFV process in type-I seesaw.} 
\label{Seesaw_I}
 \end{figure}

 In type-I seesaw, the above Hamiltonian arises due to the box diagrams shown in figure~\ref{Seesaw_I}. The internal lepton line can be either a light neutrino or a heavy neutrino. For the case of heavy neutrino exchange, the vertex factors contain light-heavy mixing matrix elements.
 Since PMNS matrix deviates from unitarity by a small amount, the GIM cancellation due to the exchange of light neutrinos is not exact. The net contribution, due to light neutrino exchange, is of the same order as the heavy neutrino exchange, as described in appendix \ref{A1}. Among the four box diagrams, the ones containing the Goldstone boson $\Phi$ have additional factors of light-heavy neutrino mixing matrix elements and are suppressed. The dominant contribution, arising from the box diagram with $2\,W$ exchange, has $(V-A)$ structure for the lepton current also, which implies that $C_{10} = - C_9$. In the approximation, where the two heavy neutrinos are nearly degenerate ($M_2 = M_1 (1+z)$,
 where $z \ll 1$), $C_9$ is given by 
 \begin{equation}
     C_{9\,(I)} \simeq x_t \left[R_{\alpha 1} R^{\ast}_{\beta 1}\right] \left[ \frac{1}{2\sin^2{\theta_W}} \,\mathcal{I}_{1} (x_t \,, x_1)\right],
     \label{C9I}
 \end{equation}
 where the function $\mathcal{I}_{1} (x_t \,, x_1)$ is defined in appendix \ref{A1}.

The scalar triplets of type-II seesaw, which mediate the radiative CLFV decays, do not couple to quarks and hence do not contribute to CLFV decays of mesons. Also, there are no heavy neutral fermions in type-II seesaw and the CLFV decays of mesons occur due to the exchange of light neutrinos only. Here, the PMNS matrix will remain unitary making the GIM suppression exact. Hence, the values of $C_9$ and $C_{10}$, in type-II seesaw, are negligibly small.  
 
\begin{figure}
    \centering
    \includegraphics[width=0.95\linewidth]{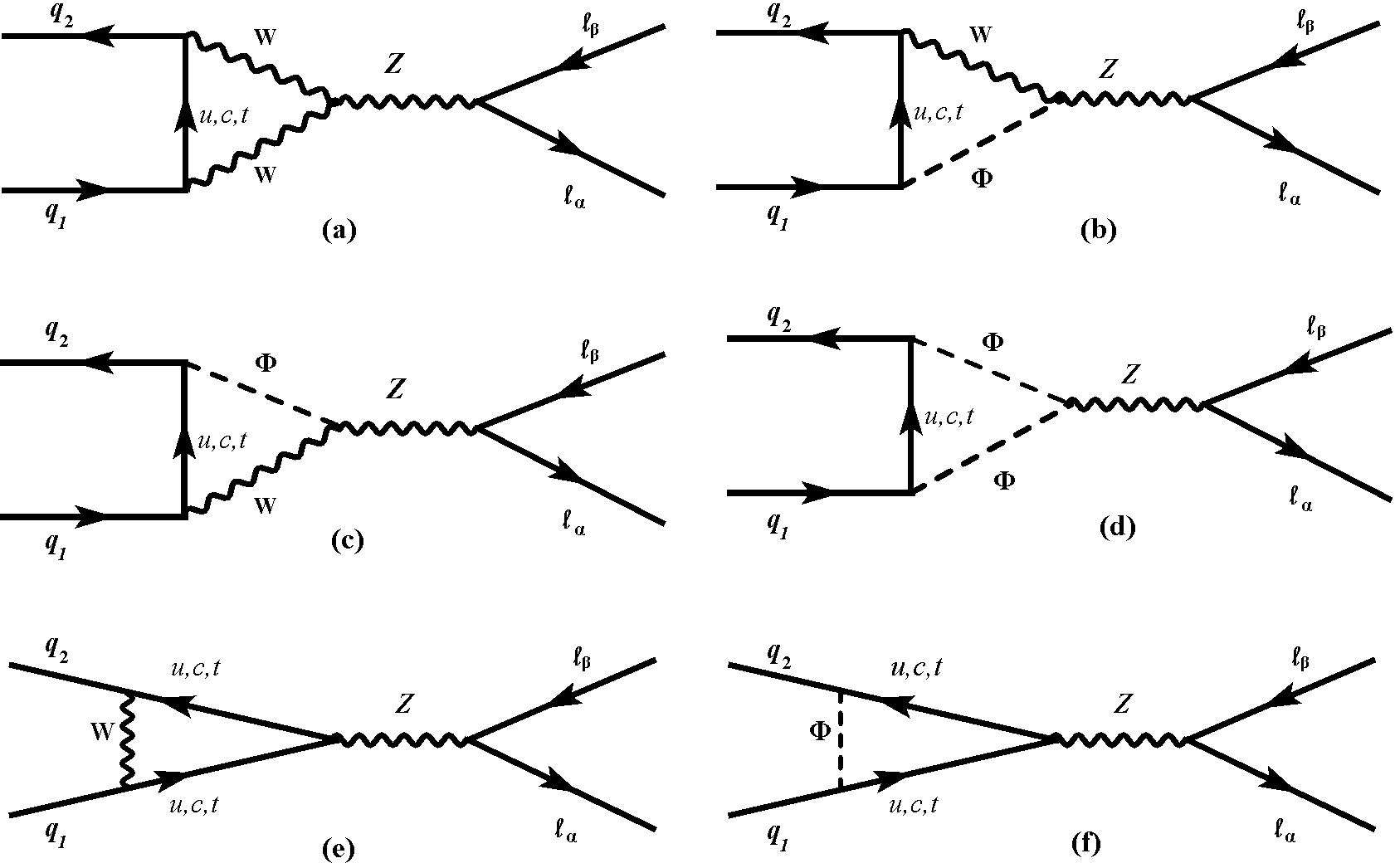}
   \caption{Additional diagrams contributing to $q_1 \to q_2 \ell_\beta^+ \ell_\alpha^-$ CLFV process in type-III seesaw.}
\label{Seesaw_III}
\end{figure}

 In the case of type-III seesaw, the box diagrams of figure~\ref{Seesaw_I} contribute to the effective Hamiltonian, where the heavy neutrino $N_k$ is replaced by the neutral component of the triplet fermion $\Sigma_k$. As in the case of type-I seesaw,
 the contribution of the box diagram with $2 \, W$ exchange dominates over the other three box diagrams. 
 However, there are additional contributions to this Hamiltonian through the triangle diagrams of 
 figure~\ref{Seesaw_III}, which occur due to the FCNC of the charged leptons to the $Z$ boson. In this case, there are no additional suppression factors for diagrams with Goldstone boson exchange. Hence, all six diagrams have comparable contributions. In principle, one can also replace $Z$ in these diagrams by the SM Higgs boson, which also has FCNC to charged leptons. However, the amplitudes of those diagrams vanish, in the limit when external fermion masses are neglected. 

The calculation of the additional diagrams for type-III seesaw is described in appendix \ref{B1}. We note that the corresponding effective Hamiltonian also has $(V-A)$ structure
for the leptonic currents, as in the case of type-I seesaw. Hence, we have $C_{10} =
- C_9$ for type-III seesaw also. Here again, we assume that the two heavy neutral fermions
belonging to the triplets, are nearly degenerate. In this approximation, we obtain 
\begin{equation}
     C_{9\,(III)} \simeq x_t \left[R_{\alpha 1} R^{\ast}_{\beta 1}\right] \left[ \mathcal{I}_{\rm tot} (x_t \,, x_1) \right],
     \label{C9III}
 \end{equation}
 where the function $\mathcal{I}_{\rm tot} (x_t \,, x_1)$ is defined in appendix \ref{B1}.
 
The effective Hamiltonian in eq.~\eqref{EFT} leads to the following CLFV decays of
pseudoscalar meson $M$:
\begin{itemize}
    \item purely leptonic decays of the form $M \to \ell_\beta^+ \ell_\alpha^-$
    \item semi-leptonic decays with a pseudoscalar meson $M^\prime$ in the final state
    $M \to M^\prime \ell_\beta^+ \ell_\alpha^-$
    \item semi-leptonic decays with a vector meson $V$ in the final state
    $M \to V \ell_\beta^+ \ell_\alpha^-$,
\end{itemize}
where we assume that $\ell_\beta$ is the more massive lepton.
The amplitudes for these decays are of the form 
\begin{eqnarray}
    T(M \to \ell_\beta^+ \ell_\alpha^-) & = & \left[ \frac{4 G_F}{\sqrt{2}} V^*_{t\,q_1} V_{t \, q_2} C_9 \right] (-i f_M) (m_{\ell_\beta}) \bar{u}_{\ell_\alpha} (1 + \gamma_5) v_{\ell_\beta}, \nonumber \\
    T(M \to M^\prime \ell_\beta^+ \ell_\alpha^-) & = & \left[ \frac{4 G_F}{\sqrt{2}} V^*_{t\,q_1} V_{t \, q_2} C_9 \right] 
     \bigg( f^+_{M M^\prime} (p_M + p_{M^\prime})_\mu + f^-_{M M^\prime} (p_M - p_{M^\prime})_\mu \bigg) \nonumber \\
    & & \bar{u}_{\ell_\alpha} \gamma^\mu (1-\gamma_5) v_{\ell_\beta}, \nonumber \\
    T(M \to V \ell_\beta^+ \ell_\alpha^-) & = & \left[ \frac{4 G_F}{\sqrt{2}} V^*_{t\,q_1} V_{t \, q_2} C_9 \right]
       \bigg\{ \varepsilon_{\mu\nu\alpha\beta}\,\varepsilon^{\ast\,\nu}\, p_M^\alpha \,p_V^\beta \frac{2V_{MV}}{m_M+m_V} \nonumber \\
        & & 
 +i(m_M+m_V)A_{MV}^1\,\varepsilon_\mu^\ast -i(\varepsilon^\ast.p_M)\bigg[\frac{A^2_{MV}}{m_M+m_V}(p_M+p_V)_\mu \nonumber \\ & & 
+\frac{2m_V}{(p_M-p_V)^2}(A^3_{MV}-A^0_{MV})(p_M-p_V)_\mu\bigg]\bigg\} \, 
        \bar{u}_{\ell_\alpha} \gamma^\mu (1-\gamma_5) v_{\ell_\beta}.\,\, 
\end{eqnarray}
In the above equations, $f_M$ is the decay constant of $M$,
$f^+_{M M^\prime}$ and $f^-_{M M^\prime}$ are the form factors for the transistion $M \to M^\prime$, $V_{M V}$, $A^i_{M V}$ (i=0,1,2,3) are the form factors for the transition $M \to V$ and $\varepsilon^\ast$ is the polarization vector of the vector meson $V$. The purely leptonic decay is subject to helicity suppression, making
the corresponding amplitude proportional to $m_{\ell_\beta}$. On the other hand, the semi-leptonic decays are free from this suppression. Hence, we expect the semi-leptonic branching ratios to be larger. 

The mixing between the light neutrinos and heavy neutral leptons, $R_{\alpha 1}$ and $ R_{\beta 1}$, 
are completely unknown in the above amplitudes, whereas the other terms are reasonably
well known. We constrain the product $| R_{\alpha 1} R^{\ast}_{\beta 1}|$ from
the experimental upper limit on the radiative decay $\ell_\beta \to \ell_\alpha \gamma$.
Substituting this constraint in the above expressions, we obtain an upper bound on $C_9$ and use it to predict upper bounds on the branching ratios of the CLFV meson decays.

\section{Upper Bounds on CLFV decays of Mesons}

\subsection{$K$ decays}
For purely leptonic kaon decay $K_L\to \mu^+ e^-$, we define $|K_L\rangle= \frac{1}{\sqrt{2}}\big(|K^0\rangle
+ |\bar{K^0}\rangle\big)$, where $CP|K^0\rangle = - |\bar{K}^0\rangle$. So, the amplitude for this decay is the sum of the amplitudes for $K^0 \to \mu^+ e^-$
and $\bar{K^0} \to \mu^+ e^-$.
The expression for this branching ratio is given by 
\begin{eqnarray}
Br(K_L \rightarrow \mu^+ e^-) = 2 \tau_{K} \frac{G^2_F \alpha^2}{32 \pi^3}f^2_{K} m_{K} m^2_{\mu} \left(1-\frac{m^2_{\mu}}{m^2_{K}} \right)^2 \left[ \Re (V_{t \, s} V^{\ast}_{t \, d}) \right]^2 |C_9|^2.
\end{eqnarray}
We evaluate 
 the branching ratio of the semi-leptonic decay $ K^+ \to \pi^+ \mu^+ e^- $ numerically, taking into account the form factors~\cite{Carrasco:2016kpy} and the three-body phase space. The Wilson coefficients satisfy the constraint $C_{10} = - C_9$   because the effective Hamiltonian in eq.~\eqref{EFT} has $(V-A)$ structure for lepton currents also for both type-I and type-III seesaw. The branching ratio for this decay is given by 
\begin{eqnarray}
Br\left(K^+ \to \pi^+ \mu^+ e^- \right) & = & a_9^K\left|C_9\right|^2,
\end{eqnarray}
where the numerical factor, $a_9^K=1.94\times 10^{-13}$, is calculated using
the form factors~\cite{Carrasco:2016kpy} and phase space integration. 

Substituting the constraint on the CLFV parameter $|R_{e1}R_{\mu 1}^\ast|$,
obtained from $\mu \to e \, \gamma$, we predict the branching ratios for the
CLFV decays of kaons. Due to the large CKM suppression arising due to $(V_{t \, s} V^{\ast}_{t \, d})$, the predicted branching ratios are quite small. In fact, they are six to nine orders of magnitude smaller than the branching ratio of $\mu \to e \, \gamma$, as listed in Table~\ref{decay}.

For meson CLFV decays, the loop functions for both type-I and type-III seesaw become independent of $M_1$ for $M_1 \gg 100$ GeV. This can be seen from the definition of $\mathcal{I}_1 (x_t, x_1)$ in eq.~\eqref{defI1} for type-I seesaw and from the definition of $\mathcal{I}_{\rm tot} (x_t,x_1)$
in eq.~\eqref{defItot} for type-III seesaw. Hence, we do not expect any $M_1$ dependent cancellations in the branching ratio of meson CLFV decays.


\subsection{$B$ decays}
The expression for the branching ratio of the purely leptonic decay of the
meson $B = (q_2 \bar{b}) \to \ell_\beta^+ \ell_\alpha^-$ is given by 
\begin{eqnarray}
Br(B  \rightarrow \ell_\beta^+ \ell_\alpha^-) = \tau_{B} \frac{G^2_F \alpha^2}{32 \pi^3}f^2_{B} m_{B} m^2_{\ell_\beta} \left(1-\frac{m^2_{\ell_\beta}}{m^2_{B}} \right)^2 |V_{t \, q_2} V^{\ast}_{t \, q_1}|^2 |C_9|^2.
\end{eqnarray}
From the definition of $C_{9(I)}$ in eq.~\eqref{C9I} and 
$C_{9(III)}$ in eq.~\eqref{C9III}, we note that these functions
are the same for both kaon decays and B meson decays.
The CLFV semi-leptonic decays of $B$ mesons are studied earlier in refs.~\cite{Becirevic:2016zri,Cirigliano:2021img, Becirevic:2024vwy}.
The expressions for the branching ratios of these decays depend on 
the form factors and the corresponding kinematic terms. In general, these 
expressions can be complicated. For the decays of $B$ mesons (both $B_d$ and $B_s$),  
ref.~\cite{Becirevic:2024vwy} provides expressions for these branching ratios in the
form of the quantities $a_9$ and $b_9$. These quantities contain the combined 
information of the form factors and phase space factors. Here again, the $(V-A)$ structure of the leptonic currents in the effective Hamiltonian implies that $C_{10} = - C_9$. With this constraint, the expressions for the branching ratios of the semi-leptonic decays of $B$ mesons, simplify to  
\begin{eqnarray}
Br\left(B \rightarrow M^\prime \ell_\beta^+ \ell_\alpha^-\right) & = & 2 a_9\left|C_9\right|^2 \times 10^{-9}, \\
Br\left(B \rightarrow V \ell_\beta^+ \ell_\alpha^-\right) & = & 2 (a_9+b_9)\left|C_9\right|^2 \times 10^{-9}.
\end{eqnarray}
The values of $a_9$ and $b_9$ are presented in table form in ref.~\cite{Becirevic:2024vwy}
for various semi-leptonic decays of $B$ mesons. 
Using the upper bounds on $C_9$ from the radiative CLFV decays, as described in section~\ref{Sec-4},
we calculate the upper bounds on the following CLFV decays of $B$ mesons:
(a) $B^+ \to K^+ \, \ell_\beta^+ \, \ell_\alpha^-$, (b) $B_d \to K^{\ast 0} \, \ell_\beta^+ \,\ell_\alpha^-$,
(c) $B_s \to \phi \, \ell_\beta^+ \, \ell_\alpha^-$, (e) $B^+ \to \pi^+ \, \ell_\beta^+ \,\ell_\alpha^-$, (e) $B_d \to \rho^0 \, \ell_\beta^+ \, \ell_\alpha^-$ and 
(f) $B_s \to \ell_\beta^+ \, \ell_\alpha^-$.

The experimental upper limits on the branching ratios of the CFLV decays of mesons are listed as the sum of the two charge conjugate modes $\ell_\beta^+ \, \ell_\alpha^-$ and $\ell_\beta^- \, \ell_\alpha^+$. Hence, our results, presented in Table~\ref{decay}, are also for the sum of 
these two modes. From the structure of our effective Hamiltonian, it is straightforward to see that the branching ratios for the two charge conjugate modes are equal. From the table, this 
following facts may be noted:
\begin{itemize}
\item The predicted upper bound for type-III seesaw is one to two orders of magnitude
larger than the one for type-I seesaw. This occurs due to the additional contributions
in type-III seesaw.
    \item The predicted upper bounds are seven to eight orders of magnitude smaller than the current upper bounds.
    \item The upper bound on semi-leptonic decay is much smaller than the corresponding radiative decay. This is due to the additional suppression from the CKM elements in the effective Hamiltonian of the semi-leptonic decay, relative to the radiative decay.
    \item The upper bound on the purely leptonic decay is even smaller compared to the
    semi-leptonic decay. This arises due to the helicity suppression of the purely leptonic
    decay amplitude. 
\end{itemize}

 \begin{table}[h!]
 \centering
\begin{tabular}{ |p{3.1cm}||p{2.5cm}||p{3.7cm}||p{3.9cm}|  }
 \hline
  Decay & Exp. Limit   & \hspace{1cm}Type-I \hspace{1.2cm} $M_1$=100\,(1000) GeV  &   \hspace{0.8cm}Type-III \hspace{0.8cm}$M_{\Sigma_1}$=100\,(1000) GeV\\
 \hline\hline
$Br(K_L\to e\mu)$\vspace{1mm} & $6.3\times 10^{-12}$~\cite{ParticleDataGroup:2024cfk} & $ 4.32 \,(1.11)\times 10^{-20}$ & $1.67\, (3.37)\times 10^{-19}$ \\
$Br(K^+\to\pi^+ e\mu)$\vspace{1mm}  &$1.1\times 10^{-10}$~\cite{ParticleDataGroup:2024cfk} & $5.48 \,(1.42)\times 10^{-22}$  & $2.12 \,(4.28)\times 10^{-21}$\\
 \hline
  $Br(B_s\to e\mu)$ & $5.4\times 10^{-9}$~\cite{LHCb:2017hag}& $2.64 \,(0.68)\times 10^{-19}$ & $1.01\,(2.06)\times 10^{-18}$\\
   $Br(B_s\to e\tau)$ & $1.4\times 10^{-3}$~\cite{Belle:2023jwr} & $4.65\, (1.20)\times 10^{-12}$ & $1.78\,(3.62)\times 10^{-11}$\\
    $Br(B_s\to\mu\tau)$ & $3.4\times 10^{-5}$~\cite{Belle:2021rod} & $ 6.19\, (0.16)\times 10^{-12}$ & $2.37\,(4.81)\times 10^{-11}$\\
 \hline
 $Br(B^+\to K^+ e\mu)$\vspace{1mm} & $1.8\times 10^{-8}$~\cite{BaBar:2007xeb}  & $1.05\,(0.27)\times 10^{-16}$ & $4.09\,(8.27)\times 10^{-16}$\\
  $Br(B^+\to K^+ e\tau)$\vspace{1mm} & $3.1\times 10^{-5}$~\cite{BaBar:2012azg} & $5.21\,(1.34)\times 10^{-12}$ & $2.00\,(4.06)\times 10^{-11}$\\
 $Br(B^+\to K^+\mu\tau)$ & $3.1\times 10^{-5}$~\cite{BaBar:2012azg} &
 $6.76\,(1.75)\times 10^{-12}$ & $2.60\,(5.25)\times 10^{-11}$\\
 \hline
 $Br(B^0\to K^{\ast 0} e\mu)$\vspace{1mm} & $1.0\times 10^{-8}$~\cite{LHCb:2022lrd} & $ 2.14\,(0.55)\times 10^{-16}$ & $0.82\,(1.67)\times 10^{-15}$\\
  $Br(B^0\to K^{\ast 0}e\tau)$\vspace{1mm} &- & $8.78\,(2.26)\times 10^{-12}$ & $3.36\,(6.84)\times 10^{-11}$\\
   $Br(B^0\to K^{\ast 0}\mu\tau)$\vspace{1mm} & $1.8\times 10^{-5}$~\cite{LHCb:2022wrs}  & $1.20\,(0.31)\times 10^{-11}$ & $4.62\,(9.36)\times 10^{-11}$\\
 \hline
  $Br(B_s\to \phi\, e\mu)$\vspace{1mm} & $1.6\times 10^{-8}$~\cite{LHCb:2022lrd} & $2.23\,(0.57)\times 10^{-16}$ & $0.86\,(1.74)\times 10^{-15}$\\
  $Br(B_s\to \phi\, e\tau)$\vspace{1mm} & - & $8.82\,(2.27)\times 10^{-12}$ & $3.37\,(6.87)\times 10^{-11}$\\
   $Br(B_s\to \phi\,\mu\tau)$\vspace{1mm} & $2.0\times 10^{-5}$~\cite{LHCb:2024wve} & $1.18\,(0.30)\times 10^{-11}$ & $4.53\,(9.18)\times 10^{-11}$\\
 \hline
 $Br(B^+\to \pi^+ e\mu)$\vspace{1mm} & $9.2\times 10^{-8}$~\cite{BaBar:2007xeb} & $3.68\,(0.95)\times 10^{-18}$ & $1.42\,(2.87)\times 10^{-17}$\\
 $Br(B^+\to \pi^+ e\tau)$\vspace{1mm} & $7.5\times 10^{-5}$~\cite{BaBar:2012azg} & $2.00\,(0.52)\times 10^{-13}$ & $0.76\,(1.56)\times 10^{-12}$\\
 $Br(B^+\to \pi^+\mu\tau)$\vspace{1mm} & $7.2\times 10^{-5}$~\cite{BaBar:2012azg} & $2.61\,(0.67)\times 10^{-13}$ & $1.00\,(2.02)\times 10^{-12}$\\
 \hline
  $Br(B^0\to \rho^0 e\mu)$\vspace{1mm} & $3.2\times 10^{-6}$~\cite{ParticleDataGroup:2024cfk} & $8.33\,(2.15)\times 10^{-18}$ & $3.22\,(6.50)\times 10^{-17}$\\
 $Br(B^0\to \rho^0 e\tau)$\vspace{1mm} &- & $3.56\,(0.92)\times 10^{-13}$ & $1.36\,(2.77)\times 10^{-12}$\\
 $Br(B^0\to \rho^0\mu\tau)$\vspace{1mm} &- & $5.28\,(1.37)\times 10^{-13}$ & $2.02\,(4.10)\times 10^{-12}$\\
 \hline
\end{tabular}
\caption{Predictions of various CLFV meson decays in seesaw models. Here we list the upper limits on decays with both combinations of lepton charges, $\ell_\alpha \ell_\beta \equiv \ell_\alpha^+\ell_\beta^- + \ell_\alpha^-\ell_\beta^+$. }
\label{decay}
\end{table}

\section{Summary and Conclusions}

Oscillation of a neutrino from one flavour to another is possible only if there
is lepton flavour violation. Any model of neutrino mass must necessarily
contain this violation. Since the light neutrinos and the charged
leptons are members of $SU(2)_L$ doublets, any flavour violation in the neutrino
sector must necessarily lead to flavour violation in the charged lepton sector also.
The amount of this CLFV depends on the details of the neutrino mass model.
It is expected that a comprehensive study of charged lepton flavour violation, 
together with neutrino oscillations, will give us important clues towards the 
construction of the correct neutrino mass model. 

Seesaw models are popular extensions of the SM, which generate tiny masses for light
neutrinos without excessive fine-tuning. In their simplest form, there are three seesaw models,
named type-I, type-II and type-III, which are distinguished by the type of heavy neutral 
particle, responsible for giving rise to light neutrino masses. There have been a number
of studies in the literature which calculated the maximum possible rates for the CLFV
radiative decays $\ell_\beta \to \ell_\alpha \, \gamma$, in these three types of seesaw. 
All these studies obtain predictions which are just below the present upper limits.

In this work, we calculate the CLFV decays of various pseudoscalar mesons in the three
types of seesaw mechanisms. The terms in the Lagrangian, which drive
the CLFV in the radiative decays, also drive the CLFV in the meson decays. However, the
relation between these two types of CLFV is different for different types of seesaw.  
Hence, the upper bound on the CLFV parameter obtained from the radiative decay, leads to
different upper bounds on the CLFV decays of the mesons in different types of seesaw.
Therefore, a comparison of the branching ratios of radiative CLFV decays and CLFV decays
of mesons can help in making a distinction between different types of seesaw. It should
be noted, however, that the upper bounds on the branching ratios of CLFV decays of mesons 
usually have much smaller than the upper bounds on the branching ratios of the radiative 
CLFV decays. This is caused by the presence of CKM matrix elements in the transition 
matrix element of the CLFV decays of mesons. 

The relation between the two kinds of CLFV decays, in each of the three types of seesaw,
can be summarised as
\begin{itemize}
    \item \underline{Type-I}: This scenario predicts the largest branching ratios of 
    semi-leptonic CLFV decays of mesons to be  
    about three orders of magnitude smaller than those of radiative CLFV decays.
    \item \underline{Type-II}: In this seesaw, the mechanism of CLFV in radiative 
    decays and that in meson decays are quite different. Hence the two branching
    ratios are not related. The branching ratio for radiative CLFV can be close to 
    the experimental upper limit~\cite{Ibarra:2011xn} but that for meson CLFV decay
    is negligibly small ($\sim 10^{-50}$) due to the exact GIM cancellation in the
    neutrino sector.
    \item \underline{Type-III}: There are additional contributions to meson CLFV decay 
    in this scenario, compared to type-I seesaw. Hence, the predicted upper bounds on
    the branching ratios for the largest of these processes are only two orders of
    magnitude smaller than those for radiative CLFV decays. 
\end{itemize}

When the CLFV decays are observed, a comparison of the branching ratios of the 
radiative decays and meson decays will enable us to identify which of the above
types of seesaw is operative in generating neutrino masses. If the branching ratio
for meson CLFV decays turn out to be larger than those of radiative CLFV decays,
we have to conclude that the origins of these two modes of CLFV are different. 
Ref.~\cite{Korrapati:2020rao} shows an example of one model where this is true. If
this indeed happens, then the mechanism of neutrino mass generation is much 
more complicated than a simple seesaw model.

\section*{Acknowledgements} 
Jai More thanks the
Department of Science and Technology (DST), Government of India for the financial support through the grant
no. SR/WOS-A/PM-6/2019(G). Purushottam Sahu and S. Uma Sankar thank the Ministry of Education, Government of India for financial support through Institute of Eminence funding to I.I.T. Bombay.

\appendix

\section{Effective Hamiltonian in Type-I seesaw}
\label{A1}

The calculation of all the diagrams in~\ref{Seesaw_I}, leads to the following effective Hamiltonian for type-I seesaw mechanism,
\begin{equation} \label{Amplitude-I}
\begin{split}
 \mathcal{H}_{eff} &= f_{I} \,\Big[\overline{q_2}\gamma^{\mu}(1-\gamma_{5}) q_1\big] \big[\bar{\ell}_\alpha \gamma^{\mu}(1-\gamma_{5}) \ell_\beta \Big] \\
 &=f_{I} \Big[\big(\overline{q_2}\,\gamma^{\mu}(1-\gamma_{5}) q_1\big) \big(\bar{\ell}_\alpha \,\gamma_{\mu}\,\ell_\beta \big) -\big(\overline{q_2}\gamma^{\mu}(1-\gamma_{5}) q_1\big) \big(\bar{\ell}_\alpha \gamma_{\mu}\gamma_5\, \,\ell_\beta \big)\Big].
  \end{split}
\end{equation}

The overall factor $f_{I}$ for type-I seesaw is given by
\begin{equation}
f_{I} =  f^{a}_{I} + f^{b+c}_{I} +f^{d}_{I}  ,
\end{equation}

where
\begin{equation} \label{BoxTI}
\begin{split}
&f^{a}_{I} =  \frac{G^2_F M^2_W}{8 \pi^2}\sum_j V^*_{j\,q_1} V_{j \, q_2} x_j \sum_k R_{\alpha \,k} R^*_{\beta \, k} \,\mathcal{I}_{1} (x_j \,, x_k), \\
&f^{b+c}_{I} =  \frac{G^2_F M^2_W}{8 \pi^2}\sum_j V^*_{j\,q_1} V_{j \, q_2} x_j \sum_k R_{\alpha \,k} R^*_{\beta \, k} \left( R_{\beta \, k} + R_{\alpha\, k}^\star \right) \, \mathcal{I}_{2} (x_j \,, x_k), \\
&f^{d}_{I} =  \frac{G^2_F M^2_W}{8 \pi^2}\sum_j V^*_{j\,q_1} V_{j \, q_2} x_j \sum_k R_{\alpha \,k} R^*_{\beta \, k} R_{\alpha \, k} R^*_{\beta \, k} \, \mathcal{I}_{2} (x_j \,, x_k) .
\end{split}
\end{equation}
The loop integral functions are
\begin{equation} \label{defI1}
\begin{split} 
& \mathcal{I}_{1} (x_j \,, x_k) =  x_k \left[ \frac{- \ln x_j }{(x_j-x_k)(1-x_j)^2}+ \frac{ \ln x_k }{(x_j-x_k)(1-x_k)^2}- \frac{1}{(1-x_j)(1-x_k)} \right], \\
& \mathcal{I}_{2} (x_j \,, x_k) =  x_k \left[ \frac{- x_j \ln x_j }{(x_j-x_k)(1-x_j)^2}+ \frac{ x_k \ln x_k }{(x_j-x_k)(1-x_k)^2}- \frac{1}{(1-x_j)(1-x_k)} \right]. 
\end{split}
\end{equation} 
By comparing the expressions given in \eqref{EFT} and \eqref{Amplitude-I}, the Wilson Coefficient $C_9$, for type-I seesaw, is obtained to be 
\begin{equation}
    C_{9\,(I)} =  C^{a}_{9\,(I)} + C^{b+c}_{9\,(I)} +C^{d}_{9\,(I)} , 
\end{equation}
where 
\begin{equation} \label{C9TI}
\begin{split} 
   & C^{a}_{9\,(I)}\simeq\frac{1}{4\sin^2{\theta_W}}\, \left(\frac{2+z}{1+z}\right) \left[R_{\alpha \, 1} R^{\ast}_{\beta \, 1}\right]\, x_t\, \mathcal{I}_{1} (x_t \,, x_1), \\
   & C^{b+c}_{9\,(I)}\simeq  \frac{1}{4\sin^2{\theta_W}}\, \left(\frac{2+z}{1+z}\right) \left[ R_{\alpha \,1} R^*_{\beta \, 1} \left( R_{\beta \, 1} + R_{\alpha\, 1}^\ast \right) \right] \, x_t\, \mathcal{I}_{2} (x_t \,, x_1), \\
    &C^{d}_{9\,(I)} \simeq\frac{1}{4\sin^2{\theta_W}}\, \left(\frac{2+z}{1+z}\right) \left[R_{\alpha \, 1} R^{\ast}_{\beta \, 1}\right]^2\, x_t\, \mathcal{I}_{2} (x_t \,, x_1). \\
\end{split}
\end{equation}
Note that $C^{b+c}_{9\,(I)}$ and $C^{d}_{9\,(I)}$ contain more number of the elements of the light-heavy mixing matrix $R$. Since these mixings are small, $C_{9\,(I)}$ is dominated by   $C^{a}_{9\,(I)}$. Neglecting $z$ (which is very small), we have 
\begin{equation} 
C^{a}_{9\,(I)}\simeq x_t \, \left[R_{\alpha 1} R^{\ast}_{\beta \, 1}\right]\, \frac{1}{2\sin^2{\theta_W}}\, \,\mathcal{I}_{1} (x_t \,, x_1).
\end{equation}

\section{Effective Hamiltonian in Type-III seesaw}\label{B1}

The effective Hamiltonian for type-III seesaw,
due to box diagrams, will have the same form as that given
in eq.~\eqref{Amplitude-I}, except that $x_k$ is denoted by 
$(M_{\Sigma_k}^2/M_W^2)$. The effective Hamiltonian from the additional diagrams of figure~\ref{Seesaw_III} is calculated to be 
\begin{equation} \label{Amplitude-III}
\begin{split}
  \mathcal{H}_{eff} &= f_{III} \,\Big[\overline{q_2}\gamma^{\mu}(1-\gamma_{5}) q_1\big] \big[\bar{\ell}_\alpha \gamma^{\mu}(1-\gamma_{5}) \ell_\beta \Big] \\
 &=f_{III} \Big[\big(\overline{q_2}\,\gamma^{\mu}(1-\gamma_{5}) q_1\big) \big(\bar{\ell}_\alpha \,\gamma_{\mu}\,\ell_\beta \big) -\big(\overline{q_2}\gamma^{\mu}(1-\gamma_{5}) q_1\big) \big(\bar{\ell}_\alpha \gamma_{\mu}\gamma_5\, \,\ell_\beta \big)\Big].
  \end{split}
\end{equation}
The overall factor $f_{III}$ for type-III seesaw is given by 
\begin{equation}
f_{III} =  f^{a}_{III} + f^{b+c}_{III} +f^{d}_{III} +f^{e}_{III}+f^{f}_{III} ,
\end{equation}
where
\begin{equation}
\begin{split}
&f^{a}_{III} =  \frac{3 G^2_F M^2_W}{4 \pi^2} \cos{\theta_W} \,\sum_k  R_{\alpha \,k} R^*_{\beta \, k}\sum_j V^*_{j q_1} V_{j q_2} \,x_j \,\mathcal{I}_{3} (x_j), \\
&f^{b+c}_{III} =  2 \frac{G^2_F M^2_W}{4 \pi^2} \sin^2{\theta_W} \,\sum_k  R_{\alpha \,k} R^*_{\beta \, k}\sum_j V^*_{j q_1} V_{j q_2} \,x_j \,\mathcal{I}_{3} (x_j),\\
&f^{d}_{III} =  \frac{ G^2_F M^2_W}{16 \pi^2} \cos{2\theta_W}\, \sum_k  R_{\alpha \,k} R^*_{\beta \, k}\sum_j V^*_{j q_1} V_{j q_2} \,x_j^2 \,\mathcal{I}_{3} (x_j),\\
&f^{e}_{III} =\frac{G^2_F M^2_W}{4 \pi^2} \sum_k  R_{\alpha \,k} R^*_{\beta \, k}\sum_j V^*_{j q_1} V_{j q_2}x_j\left[ (1-\frac{4}{3}\sin^2{\theta_W})\,\mathcal{I}_4 (x_j)-\frac{4}{3}\sin^2{\theta_W}\,\mathcal{I}_5 (x_j)
\right],\\
&f^{f}_{III} =\frac{G^2_F M^2_W}{8 \pi^2} \sum_k  R_{\alpha \,k} R^*_{\beta \, k}\sum_j V^*_{j q_1} V_{j q_2}x_j^2\left[ \frac{4}{3}\sin^2{\theta_W}\,\mathcal{I}_4 (x_j)-(1-\frac{4}{3}\sin^2{\theta_W})\,\mathcal{I}_5 (x_j)
\right].
\end{split}
\end{equation}
The loop integral functions are given by
\begin{equation}
\begin{split}
   & \mathcal{I}_{3} (x_j) =\frac{1-x_j+x_j\ln{x_j}}{(1-x_j)^2},\\
    &\mathcal{I}_{4} (x_j)=\frac{x_j-1-\ln{x_j}}{(1-x_j)^2}+\frac{1-x_j+x_j\ln{x_j}}{2(1-x_j)^2},\\
  &  \mathcal{I}_{5} (x_j)=\frac{x_j-1-\ln{x_j}}{(1-x_j)^2}.
    \end{split}
\end{equation}

The Wilson Coefficient for type-III seesaw is
\begin{equation}
    C_{9\,(III)} =  C^{\rm box}_{9\,(III)} + C^{a}_{9\,(III)} + C^{b+c}_{9\,(III)} +C^{d}_{9\,(III)}+C^{e}_{9\,(III)}+C^{f}_{9\,(III)},
\end{equation}
where
\begin{equation}
\begin{split}
&C^{\rm box}_{9 \,(III)} \simeq  \frac{1}{4 \sin^2{\theta_W}} \,\left(\frac{2+z}{1+z}\right) \left[R_{\alpha 1} R^{\ast}_{\beta 1}\right]\, \,x_t \,\mathcal{I}_{1} (x_t \,, x_1) ,\\
&C^{a}_{9 \,(III)} \simeq  \frac{3 \cos{\theta_W} }{2\sin^2{\theta_W}} \,  \left(\frac{2+z}{1+z}\right) \left[R_{\alpha 1} R^{\ast}_{\beta 1}\right] \,x_t \,\mathcal{I}_{3} (x_t), \\
&C^{b+c}_{9\,(III)} \simeq 2 \, \left(\frac{2+z}{1+z}\right) \frac{1}{2}\left[R_{\alpha 1} R^{\ast}_{\beta 1}\right] \,x_t \,\mathcal{I}_{3} (x_t),\\
&C^{d}_{9 \,(III)} \simeq  \frac{\cos{2\theta_W}}{8 \sin^2{\theta_W}}\, \left(\frac{2+z}{1+z}\right) \left[R_{\alpha 1} R^{\ast}_{\beta 1}\right] \,x_t^2 \,\mathcal{I}_{3} (x_t),\\
&C^{e}_{9 \,(III)} \simeq\frac{1}{ 2\sin^2{\theta_W}}\, \left(\frac{2+z}{1+z}\right) \left[R_{\alpha 1} R^{\ast}_{\beta 1}\right]\,x_t\,\left[ \left(1-\frac{4}{3}\sin^2{\theta_W}\right)\,\mathcal{I}_4 (x_t)-\frac{4}{3}\sin^2{\theta_W}\,\mathcal{I}_5 (x_t)
\right],\\
&C^{f}_{9 \,(III)} \simeq\frac{1}{4\sin^2{\theta_W}} \,\left(\frac{2+z}{1+z}\right) \left[R_{\alpha 1} R^{\ast}_{\beta 1}\right]\,x_t^2\,\left[ \frac{4}{3}\sin^2{\theta_W}\,\mathcal{I}_4 (x_t)-\left(1-\frac{4}{3}\sin^2{\theta_W}\right)\,\mathcal{I}_5 (x_t)
\right].
\end{split}
\end{equation}
In the limit $z \ll 1$, the expression for
$C_9$ in type-III seesaw is 
\begin{equation} 
C^{a}_{9\, (III)}\simeq x_t \, \left[R_{\alpha \, 1} R^{\ast}_{\beta \, 1}\right]\, \mathcal{I}_{\rm tot} (x_t \,, x_1),
\end{equation}
where
\begin{equation} \label{defItot}
\begin{split}
    \mathcal{I}_{\rm tot}(x_t \,, x_1) = &\frac{1}{2 \sin^2{\theta_W}}\,\left[\mathcal{I}_{1} (x_t \,, x_1) + \left(6 \cos{\theta_W}+
     2 +\frac{1}{ 2}\cos{2\theta_W}\,x_t\right)\mathcal{I}_{3} (x_t) \right]\\ &
+
\frac{1}{ \sin^2{\theta_W}}\,\left[ \left(1-\frac{4}{3}\sin^2{\theta_W}\right)\,\mathcal{I}_4 (x_t)-\frac{4}{3}\sin^2{\theta_W}\,\mathcal{I}_5 (x_t)
\right]  \\ & +
\frac{1}{2\sin^2{\theta_W}} \,x_t\,\left[ \frac{4}{3}\sin^2{\theta_W}\,\mathcal{I}_4 (x_t)-\left(1-\frac{4}{3}\sin^2{\theta_W}\right)\,\mathcal{I}_5 (x_t)
\right].
     \end{split}
\end{equation}

\clearpage
\bibliographystyle{utcaps_mod}
\bibliography{seesaw_ref}
\end{document}